\documentclass[twocolumn,showpacs,amsmath, preprintnumbers,amssymb]{revtex4-1}
\usepackage{graphicx}
\usepackage{dcolumn}
\usepackage{bm}

\begin{document}

\preprint{deposit manuscript by VINITI, 03.05.2012, N204-Â2012(in Russia)}

\title{Casimir force of expulsion}

\author{Evgeny\,G.\,Fateev}

 \email[E-mail: ]{e.g.fateev@gmail.com}
\affiliation{%
Institute of mechanics, Ural Branch of the RAS, Izhevsk 426067, Russia
}%
\date{\today}
\hyphenation{title}
\begin{abstract}
The possibility in principle is shown that the noncompensated Casimir force 
can exist in nanosized open metal cavities. The force shows up as 
time-constant expulsion of open cavities toward their least opening. The 
optimal parameters of the angles of the opening, of ``generating lines'' of 
cavities and their lengths are found at which the expulsive force is 
maximal. The theory is created for trapezoid configurations, in particular 
for parallel mirrors which experience both the transverse Casimir pressure 
and longitudinal compression at zero general expulsive force.
\end{abstract}

\maketitle

To find such geometrical configurations of nanosized bodies, at which the 
Casimir repulsion of the bodies, often associated with levitation and 
buoyancy effects, can take place, is an urgent task (see, for example, 
~\cite {Leonhardt:2007, Levin:2010, Rahi:2011}. 
However, the levitation is possible only at limited 
distances from the surface of bodies-partners. Much more promising and 
important is the search of geometrical shapes which can have Casimir 
noncompensated force causing the effect of continuous expulsion of a 
configuration in a certain direction independent of the proximity to the 
surfaces of bodies-partners. The objective of the present work is to 
demonstrate such possibility. 

When the volume of a quantum field is bounded by material boundaries, there 
can be not only Casimir attraction force~\cite{Casimir:1948}
but the effects of neutral ``buoyancy'' of curved 
surfaces \cite{Jaffe:2005} and other interesting phenomena as well (see, 
for example, ~\cite{Bordag:2001, Rodriguez:2011}). 
The Casimir classical result for two planes is 
obtained by assuming that they are infinite and, correspondingly, all the 
other components of the energy-momentum tensor (EMT) of the electromagnetic 
field are compensated except those that are normal to the surface. 
Naturally, in the case with objects having finite sizes the application of 
the classical Casimir result cannot reflect all possible occurrences of zero 
electromagnetic-field oscillations. The calculations of Casimir forces for 
particular configurations are rather complicated \cite{Brown:1969}. 
Therefore, light variants of approximated calculations are used taking into 
account only the existence of EMT components normal to a surface. Such 
approximation is the one of Deryagin~\cite{Derjaguin:1956}
for short-range Van der Waals forces, within the 
frames of which the Casimir forces between microspheres and a surface (see, 
for example, \cite{Bordag:2006}) and some other similar configurations 
are calculated. The discrepancies between the magnitudes of forces 
calculated with the use of the proximity force approximation (PFA) and true 
magnitudes of Casimir forces can be much more than 0.1{\%} 
\cite{Gies:2006} since there is an essential difference in the nature and 
long-range action of Casimir and Van der Waals forces. At present new 
methods for the calculation of forces for different configurations are being 
looked for (see, for example, ~\cite{Bordag:2009, Pavlovsky:2011, Jaffe:2004, Graham:2010,
Rahi:2012, Zaheer:2010}). The method based on classical geometrical optics can 
be mentioned as one of the most promising ~\cite{Jaffe:2004, Scardicchio:2005}.
However, none of the existing 
approaches allows to demonstrate the existence of Casimir expulsive forces 
in any configurations. It is due to the fact that normally problems are 
focused upon the interaction of parts of the configuration of a body or 
bodies-partners. 

Let us consider a configuration with a trapezoid cavity which will serve as 
an example for the demonstration of the possibility in principle that the 
Casimir expulsive forces can exist. Cavity should be understood as an open 
thin-walled metal shell with one or several outlets. The inner and outer 
surfaces of the cavity should have the properties of perfect mirrors. The 
cavity should entirely be immerged into a material medium or be a part of 
the medium with the parameters of dielectric permeability being different 
from those of physical vacuum. In Cartesian coordinates the configuration 
looks like two thin metal plates with the surface width $L$ (oriented along the 
$z$-axis) and length $R$, which are situated at a distance $a$ from one another; the 
angle $2\varphi$ of the opening of the generating lines of cavities between the 
plates can be varied (by the same value $\varphi $ simultaneously for both 
wings of the trapezoid cavity) as it is shown in ~Fig.\hyperlink{fig1}{}1. 
\begin{figure}
\hypertarget{fig1}{}
\centerline{\includegraphics[width=2.67in,height=2.14in]{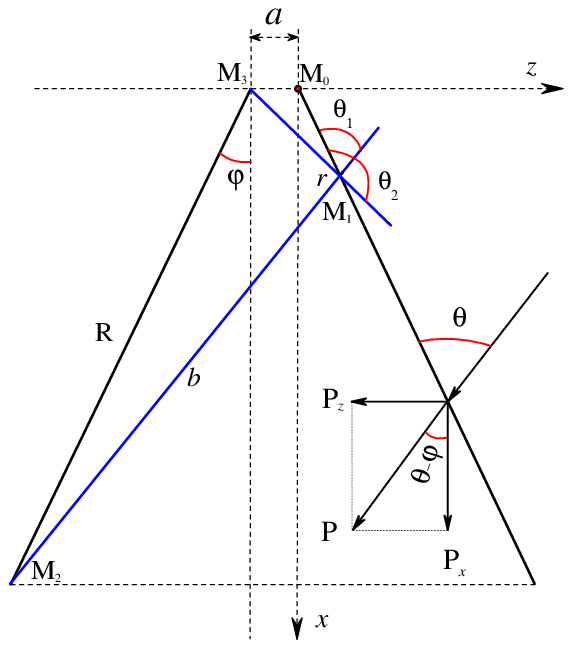}}
\caption{Schematic view of the configuration of a symmetric trapezoid cavity 
with the length $R$ of the wing surface, a particular case of which is parallel 
planes, i.e. $\varphi =0$, and a triangle at $a=0$. The section of the cavity 
shown in the Cartesian coordinates in the plane ($x,z$) has the width $L$ in the 
$y$ direction normal to the plane of the figure. The blue straight lines 
designate virtual rays with the length $b$ coming from the point $\mbox{M}_1 $ at 
limit angles $\Theta _1 $ and $\Theta _2 $ onto the right cavity surface 
ending at the ends of the opposite cavity wing at the points $\mbox{M}_2 $ 
and $\mbox{M}_3$, respectively. }
\end{figure}

Further, let us use the following formalism. Let us take the so-called 
tensor of Casimir stress for the electromagnetic field between two parallel 
plates separated by the distance $a$, which is determined in 4D space-time 
coordinates $(tc,\hat {x},\hat {y},\hat {z})$ \cite{Brown:1969, Fulling:2007}
in the form 
\begin{equation}
\label{eq1}
T^{^{\mu \nu }}=\frac{E_c }{a}\mbox{diag}(1,-1,-1,3),
\end{equation}
where the Casimir energy per unit of area is 
$E_c =-\frac{\hbar c \pi ^2}{720{\kern 1pt}a^3}.$
Here, $\hbar =h/2\pi $ is the reduced Planck constant, $c$ is the velocity of 
light. We use only the component $P_c =T^{33}$ of the stress tensor for each 
point $r$ on the cavity surface. However, let us note that the coordinates 
$(\hat {x},\hat {y},\hat {z})$ in 4D space are not the same as the 
coordinates on our drawing plane $(x,y,z)$. 

For solving the problem let us assume that virtual photons have the 
properties of rays and are strictly specularly reflected from the perfectly 
conducting metal walls both inside and outside the cavities. As a first 
approximation, we assume that this property of rays is characteristic of all 
lengths of waves of the electromagnetic spectrum. For each ray incident at 
the angle $\Theta $ on the cavity surface from the outside there is a 
strictly opposite corresponding ray coming from the inside. The action of 
the pulses of the above rays results in the presence of the local (at the 
point $r$) specific Casimir force which is noncompensated for each direction 
$\Theta$ on the outer surface of the cavity. It means that the Casimir 
pressure of each individual ray incident on the point $r$ at any of the angles 
$\Theta$ will correspond to the value 
\begin{equation}
\label{eq3}
P_c (a,\varphi ,\Theta ,r)=-\frac{\hbar c \pi ^2}{240\,b^4},
\end{equation}
where $b=f(r,a,\varphi ,\Theta ).$
The force action of each ray in our coordinates $(x,y,z)$ is decomposed into 
components directed along $x-$ and $z-$axes. The component directed against the 
$z$-axis is associated with the specific force (pressure) acting upon the 
cavity surface and its sign conforms to the sign of the Casimir pressure in 
formula (\ref{eq3}). The component directed along the $x$-axis (its sign should be +) 
is associated with the specific force of expulsion of the entire 
configuration (or the force of sliding of one cavity wing relative to the 
other when one plate is fixed and the other is not). 

The material medium both inside and outside metal cavities, in principle, 
can have dielectric permeability $\varepsilon $ different from that of 
vacuum. This fact like many others (temperature \cite{Brown:1969}, 
etc.) can be taken into account; however, further we shall consider an ideal 
situation. 

The operator of turning of the right wing of the configuration through the 
angle $\varphi$ and of each ray at any local point $r$ (with the angles $\Theta$  
 between the rays and the cavity surface as in Fig.\hyperlink{fig1}{1}) can be written in the form
\begin{equation}
\label{eq4}
\phi ^i=\left[ {{\begin{array}{*{20}c}
 {\cos (\Theta -\varphi )} \hfill \\
 {\sin (\Theta -\varphi )} \hfill \\
\end{array} }} \right].
\end{equation}
The local stress tensor at the point $r$ in our coordinates $(x,z)$ can be 
expressed as
\begin{equation}
\label{eq5}
P_{ik} =P_c \left[ {{\begin{array}{*{20}c}
 {-1} \hfill & 0 \hfill \\
 0 \hfill & 1 \hfill \\
\end{array} }} \right].
\end{equation}
Then the integral quantities for Casimir pressures along the $x-$ and $z$-axes at 
the point $r$ at all possible angles $\Theta $ are
\begin{equation}
\label{eq6}
P_k (r)=\int\limits_{\Theta _1 }^{\Theta _2 } P_{ik} \phi ^id\Theta .
\end{equation}
The entire Casimir force of compression (at $P_z =P_{22} )$ acting thus upon 
one of the plates along the $z$-axis in our coordinates $(x,y,z)$ can be 
expressed as follows
\begin{equation}
\label{eq6a}
F_z =\int\limits_0^L {dy} \int\limits_0^R {dr} \int\limits_{\Theta _1 
}^{\Theta _2 } P_c (\varphi ,\Theta ,r)\sin (\Theta -\varphi )d\Theta .
\end{equation}
The entire Casimir force of expulsion (at $P_x =P_{11} )$ along the $x$-axis is 
\begin{equation}
\label{eq7}
F_x =-\int\limits_0^L {dy} \int\limits_0^R {dr} \int\limits_{\Theta _1 
}^{\Theta _2 } P_c (\varphi ,\Theta ,r)\cos (\Theta -\varphi )d\Theta .
\end{equation}
Here for the given geometry the limit angles are $\Theta _1 =f(r,a,\varphi 
)\;$and $\Theta _2 =f(r,a,\varphi )$. Let us find these angles using geometric 
notion of directing vectors corresponding to the rays for limit angles 
$\Theta _1 ,\;\Theta _2 $ and to the ''generatrix'' of the right wing 
of the cavity on the schematic view of the trapezoid configuration (Fig.\hyperlink{fig1}{1}). 
Let us designate the point data on the trapezoid cavity scheme: $M_0 (x_0 
,z_0 )$, $M_1 (x_1 ,z_1 )$, $M_2 (x_2 ,z_2 )$ and $M_3 (x_3 ,z_3 )$. Then, 
the vector $\overrightarrow {M_0 M} _1 =(x_1 -x_0 ;z_1 -z_0 )$ is chosen as 
a directing vector of the straight line of the ``generatrix'' of the cavity 
right wing. $\overrightarrow {M_1 M} _2 =(x_2 -x_1 ;z_2 -z_1 )$ is chosen as 
a directing vector of the ray $b$. Let us choose $\overrightarrow {M_1 M} _3 
=(x_3 -x_1 ;z_3 -z_1 )$ as a vector of the ray between the extreme upper 
point of the left wing $M_3 (x_3 ,z_3 )$ and the point $M_1 (x_1 ,z_1 )$. 
The corresponding coordinates can be written in the form: $x_0 =0$; $z_0 
=0$; $x_1 =r\cos \varphi $; $z_1 =r\sin \varphi $; $x_2 =R\cos \varphi $; $z_2 
=-R\sin \varphi -a$; $x_3 =0$; $z_3 =-a$. The cosines of the angles between the 
guiding lines are defined as
$\cos \Theta _1 = \frac{\overrightarrow {M_0 M} _1 \cdot \overrightarrow 
{M_1 M} _2 }{\left\| {\overrightarrow {M_0 M} _1 } \right\|\cdot \left\| 
{\overrightarrow {M_1 M} _2 } \right\|}\nonumber $
and
$\cos {\Theta _2} =  \frac{{{{\overrightarrow {{M_0}M} }_1} \cdot {{\overrightarrow {{M_1}M} }_3}}}{{\left\| {{{\overrightarrow {{M_0}M} }_1}} \right\| \cdot \left\| {{{\overrightarrow {{M_1}M} }_3}} \right\|}}.$
Thus, the expressions for $\Theta _1 ,\;\Theta _2 $ and $b$ have the 
forms
 \begin{center}
$\Theta _1 =\mbox{arccos}\left( {-\frac{r+a\sin \varphi -R\cos 2\varphi }{\sqrt 
{\left( {a+R\sin \varphi +r\sin \varphi } \right)^2+\left( {r\cos \varphi -R\cos \varphi 
} \right)^2} }} \right),$
$\Theta _2 =\mbox{arccos}\left( {-\frac{r+a\sin \varphi }{\sqrt {a^2+r^2+2ra\sin 
\varphi } }} \right),$
$b=\frac{\sin (2\varphi -\Theta _2 )(a+r\sin \varphi )}{\sin (\Theta -2\varphi )\sin 
(\varphi -\Theta _2 )}.$
\end{center}

Let us consider the nature of the non-compensation of the forces along the 
$x$-axis. The integration (\ref{eq6}) by all the angles $\Theta$ for local specific 
forces of compression in the $z$ direction gives 
\begin{equation}
\label{eq8}
\begin{gathered}
  {P_z}(r) =  - \frac{{\hbar c{\pi ^2}}}{{{{240}^{}}{s^4}}}\int\limits_{{\Theta _1}}^{{\Theta _2}} {} \sin {(\Theta  - 2\varphi )^4}\sin (\Theta  - \varphi )d\Theta,  
\end{gathered} 
\end{equation}
where
$s=\frac{\sin (2\varphi -\Theta _2 )(a+r\sin \varphi )}{\sin (\varphi -\Theta _2 )}.$
By integrating (\ref{eq6}) by all the angles $\Theta$ for local specific forces of 
expulsion in the $x$ direction we obtain
\begin{equation}
\label{eq9}
\begin{gathered}
  {P_x}(r) = \frac{{\hbar c{\pi ^2}}}{{{{240}^{}}{s^4}}}\int\limits_{{\Theta _1}}^{{\Theta _2}} {} \sin {(\Theta  - 2\varphi )^4}\cos (\Theta  - \varphi )d\Theta.
\end{gathered}
\end{equation}
We should note that when the actions at the angles $\Theta$ are taken into 
account, for parallel plates ($\varphi =0$) at $R\to \infty $ the pressure 
$P_z(R)$ is larger than that calculated by Casimir in ${16}/{15}$ by a factor of 
$\mathop {\lim }\limits_{\varphi \to 0,\;R\to \infty } P_z (R)\to -\frac{\hbar 
c\pi ^2}{240\,s^4}\frac{16}{15}$ according to expression (\ref{eq8}). 

In this case, for the same limits, specific forces of expulsion 
$P_x(r)$ tend to zero as it should hold in the given configuration
$\mathop {\lim }\limits_{\varphi \to 0,\;\,R\to \infty } P_x (R)\to 0.$
However, the solution of (\ref{eq9}) in the limits from $\Theta _1 =0$ to $\Theta 
_2 =\pi /2$ and, separately, from $\Theta _1 =\pi /2$ to $\Theta _2 =\pi $ 
gives
\begin{equation}
\label{eq10}
P_x (\varphi \to 0,R\to \infty )=\frac{\hbar c\pi ^2}{240 s^4}\left\{ 
{\begin{array}{l}
-\frac{1}{5};\; 0\leq \Theta \leq \frac{\pi }{2} \\ 
+\frac{1}{5}; \frac{\pi}{2}\leq \Theta \leq \pi. \\ 
 \end{array}} \right.
\end{equation}
It means that the configuration of two parallel plates shrinks both along 
the $z$-axis and the $x$-axis with the specific force of the order of 1/5 of the 
classical Casimir specific force and 16/75 of that determined by formula (\ref{eq8}).

Thus, here we have a basic scheme of the calculation of the Casimir forces 
of pressure and expulsion in geometric approximation. Perhaps, it is 
possible to construct the theory of Casimir forces of expulsion based on 
first principles. However, the solutions of these and even more traditional 
problems encounter essential difficulties. Known attempts to calculate 
Casimir pressure dependences in wedged geometries with angle $2\varphi $ 
between the surfaces of the cavity lead to the results which are far from 
being real \cite{Deutsch:1979, Dowker:1978}. 

The use of formulae (\ref{eq5}-\ref{eq9}) allows to reveal the following character of 
Casimir specific forces $P_z (r)$ and $P_x (r)$ along the $x$-axis for two 
parallel plates ($\varphi =0$) of the same length $R$ with the distance between 
them $a=4\times 10^{-7}$ \,m (see Fig.~\hyperlink{fig2}{2}). The same character of the dependences 
will take place at rescaling the dimensional parameters of a configuration 
to any small values but, naturally, within the frames of physically 
reasonable limits restricted by sizes of atoms.
\begin{figure}
\hypertarget{fig2}{}
 \includegraphics[width=0.4\linewidth]{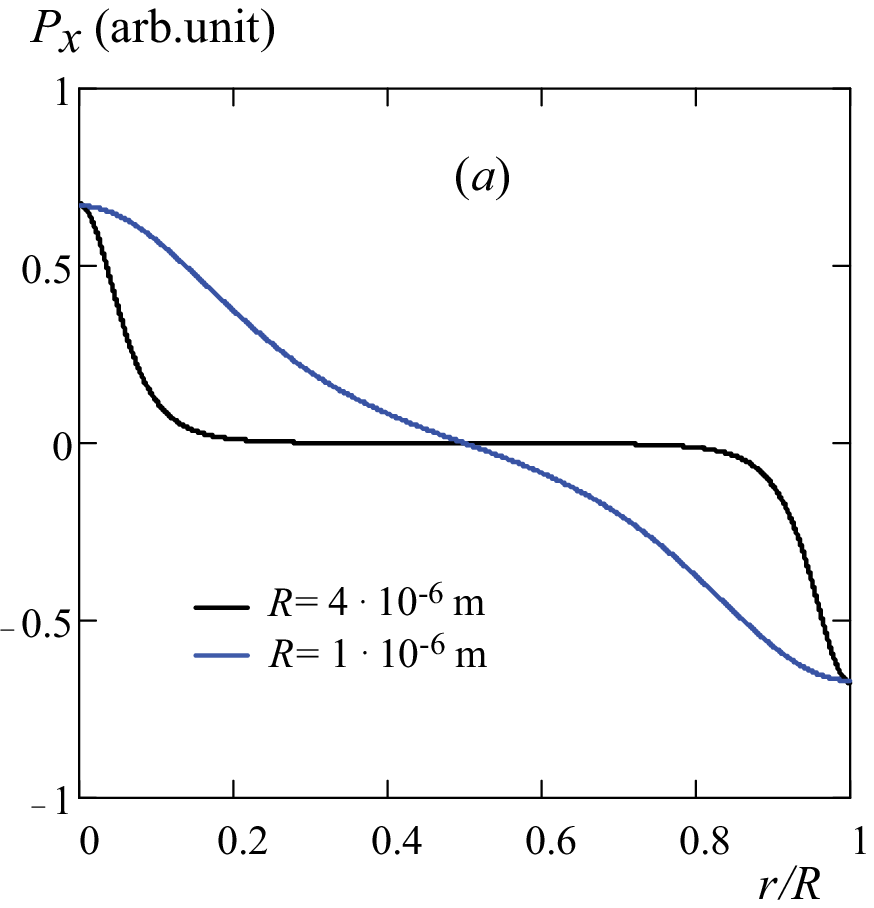}
 \hspace{0.1\linewidth}
 \includegraphics[width=0.4\linewidth]{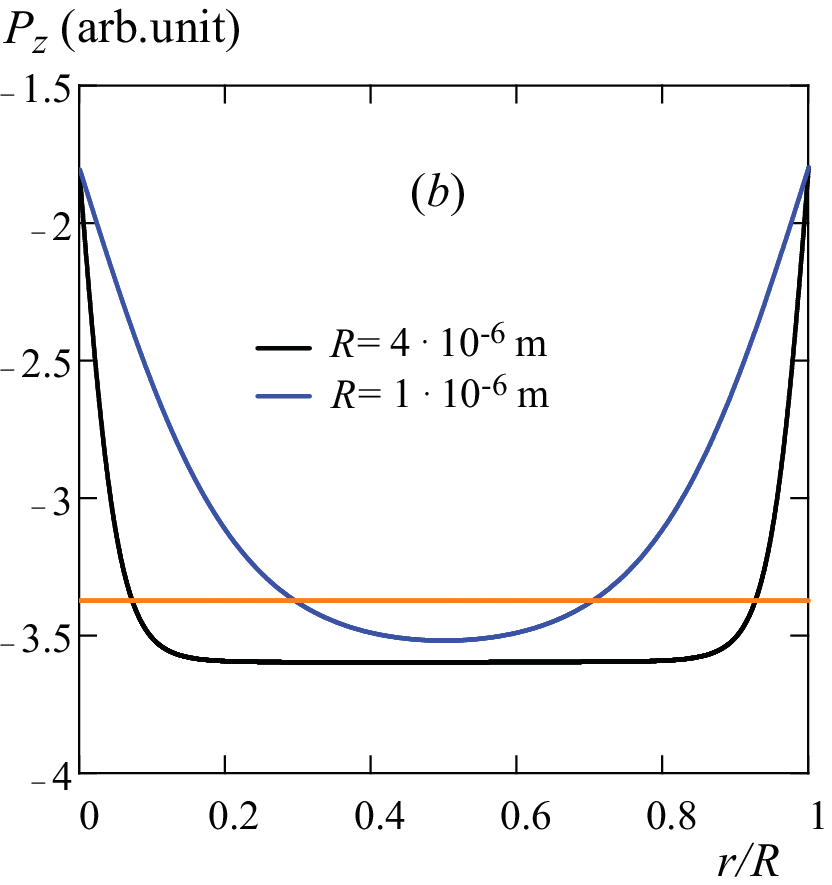}
 \caption{Specific local Casimir forces of expulsion ($a$) and pressure ($b$) along 
the $x$-axis at $\varphi =0$. The red line ($b$) shows the classical level of the 
Casimir pressure.} 
\end{figure}
In Fig.~\hyperlink{fig2}{2$b$} it can be seen that at $R/a\geqslant 1$ the specific force $P_z 
(r)$ on the boundaries is always half of that in the centre of the 
configuration. The smaller are the configuration sizes, the more uniform are 
the compression forces. In addition, it can be seen that in the 
configuration there are forces $P_x (r)$ compressing the ends of the 
parallel plates toward their centre (Fig. \hyperlink{fig2}{2$a$}). The specific force of 
compression is $16/75$ of the specific force of pressure $P_z (r)$ on the 
plates. In this case, the smaller is the length $R$, the larger part of the 
cavity wing is subjected to the action of such forces. However, along the 
$x$-axis, the integral Casimir forces $P_x (r)$ compensate one another. 

An essentially different situation can be observed when we start changing 
the angle $\varphi $ which is half of the angle between the cavity surfaces. It 
can be seen in Fig. \hyperlink{fig3}{3} that at all the angles $\varphi \ne 0$ there is a 
noncompensated force $P_x (r)$ pushing the figure along the $x$-axis both in the 
positive and negative directions. The result of the action of the 
noncompensated force will be called Casimir expulsion. 
\begin{figure}
\begin{center}
 \hypertarget{fig3}{}
 \includegraphics[width=0.425\linewidth]{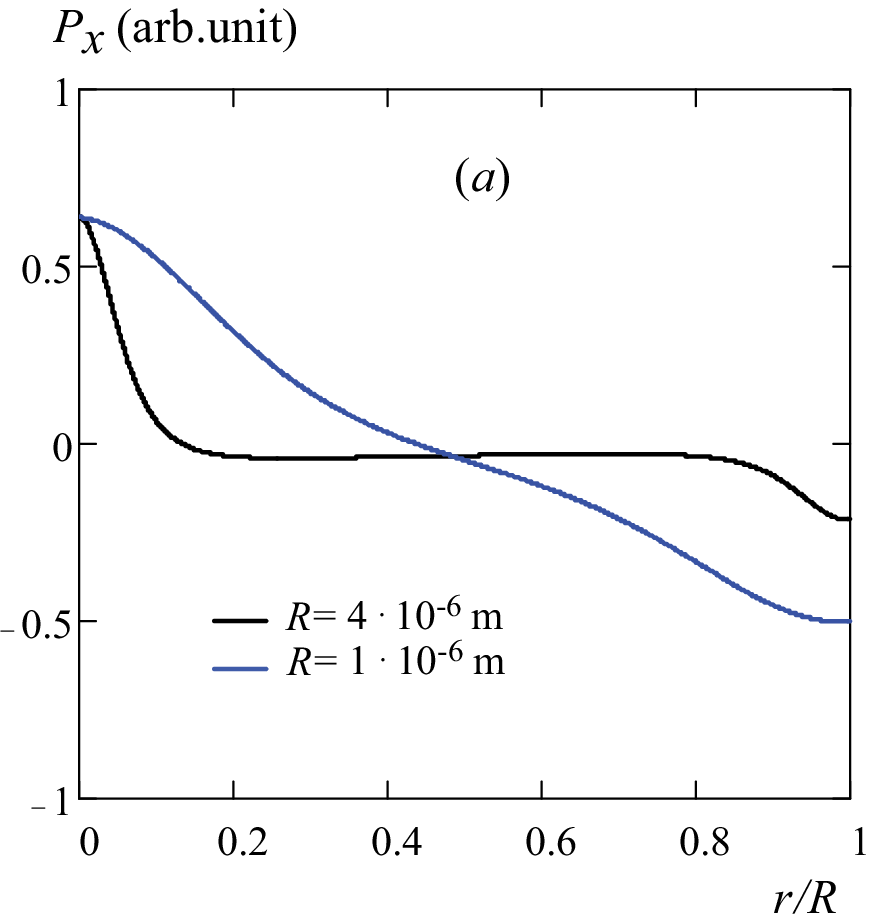}
 \hspace{0.1\linewidth}
 \includegraphics[width=0.4\linewidth]{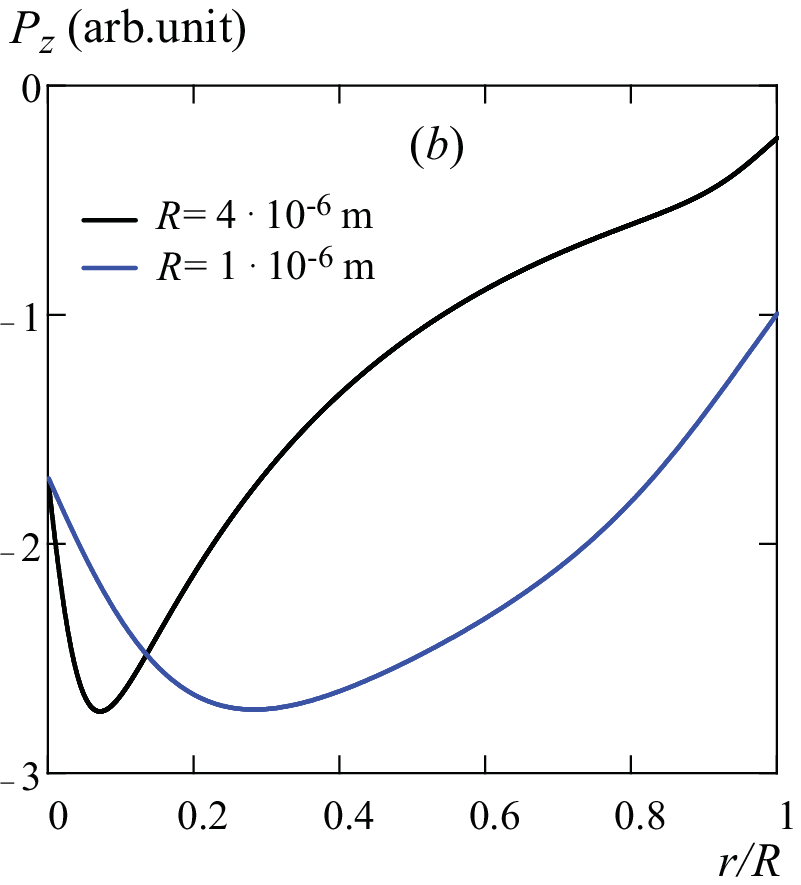}
 \end{center}
  \caption{Specific local Casimir forces of expulsion and compression along the 
$x$-axis at the angles $\varphi =1\mathring { }$ ($a,b$).}
\end{figure}
Along the $x$-axis locally near the narrower section of the trapezoid cavity 
the force pushing the cavity in the $x$ direction prevails at small angles. 
However, at the further growth of the angle $\varphi $ in the proximity of the 
narrow section of the cavity the forces pushing it in the opposite direction 
to the $x$-axis are growing. 

Having ultimately integrated $P_x (r)$ and $P_z (r)$ by $x$ we will find the 
entire Casimir force $F_x $ and $F_z $ acting upon one wing of the cavity. 
Let us study its dependence on the wing length $R$. The corresponding results are 
displayed in Fig. \hyperlink{fig4}{4}. 
\begin{figure}
\begin{center}
 \hypertarget{fig4}{}
 \includegraphics[width=0.4\linewidth]{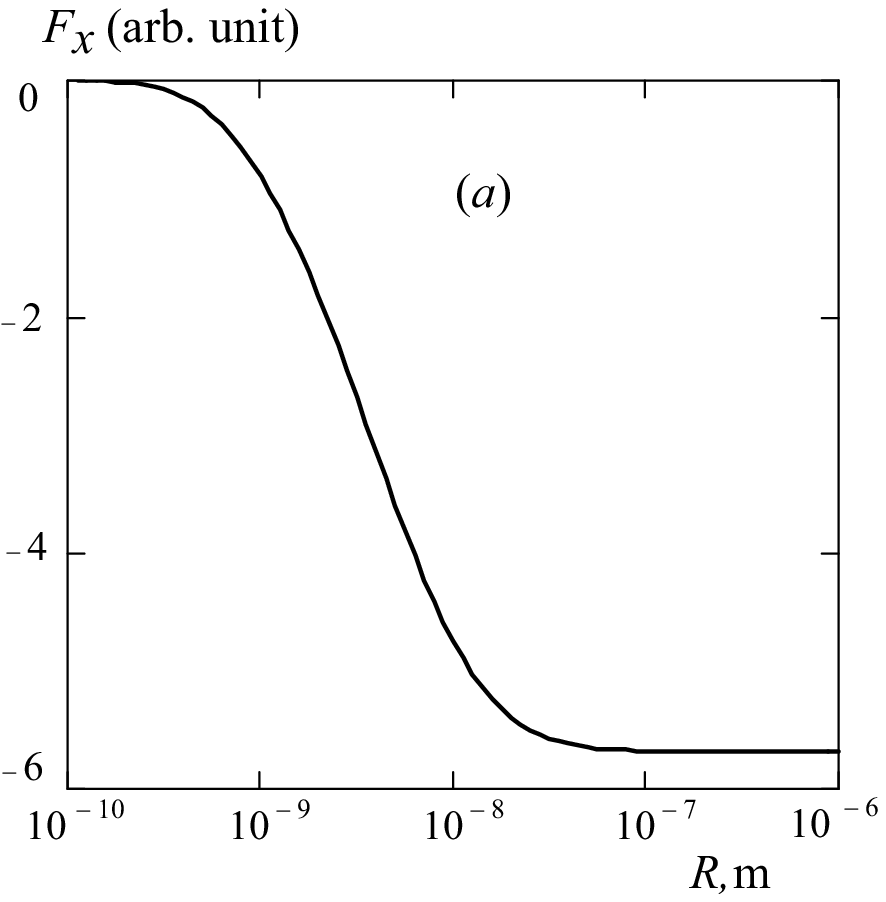}
 \hspace{0.1\linewidth}
 \includegraphics[width=0.425\linewidth]{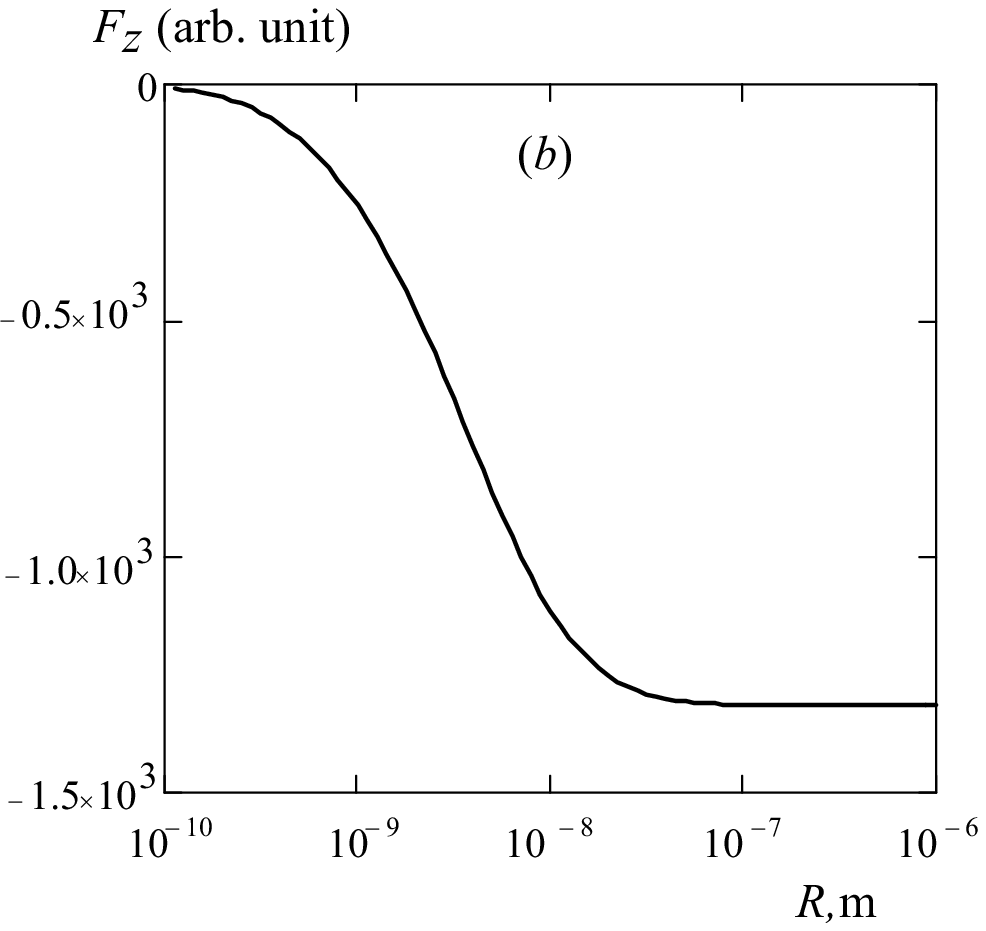}
\end{center}
\caption{Entire Casimir force of expulsion ($a$) and compression ($b$) of the 
cavity wing at $\varphi =1\mathring{ }$ depending on the wing length $R$ and width 
$L$ = 1 \, m.  }
\end{figure}
As seen from Fig.\hyperlink{fig4}{4$a$}, the total force of expulsion is 
always oppositely directed to the $x$-axis. The force appears as time-constant 
expulsion of the open trapezoid cavity in the direction of its least opening 
(i.e. in the direction of smaller section).  
\begin{figure}
\begin{center}
 \hypertarget{fig5}{}
\includegraphics[width=0.45\linewidth]{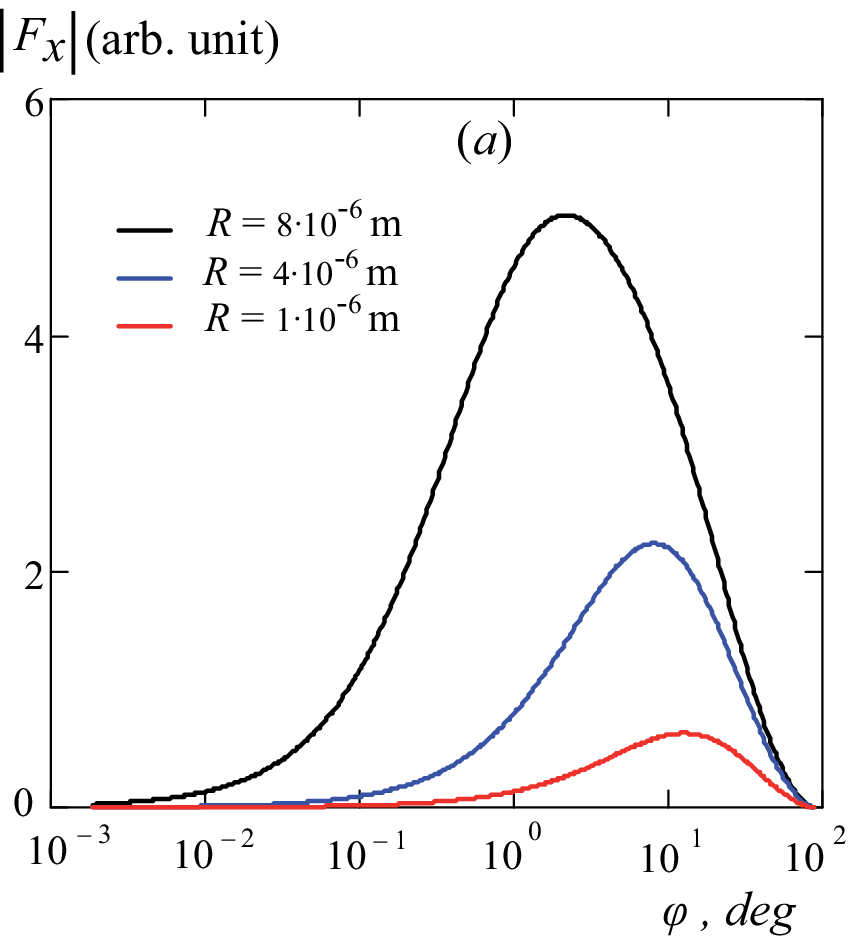}
\includegraphics[width=0.45\linewidth]{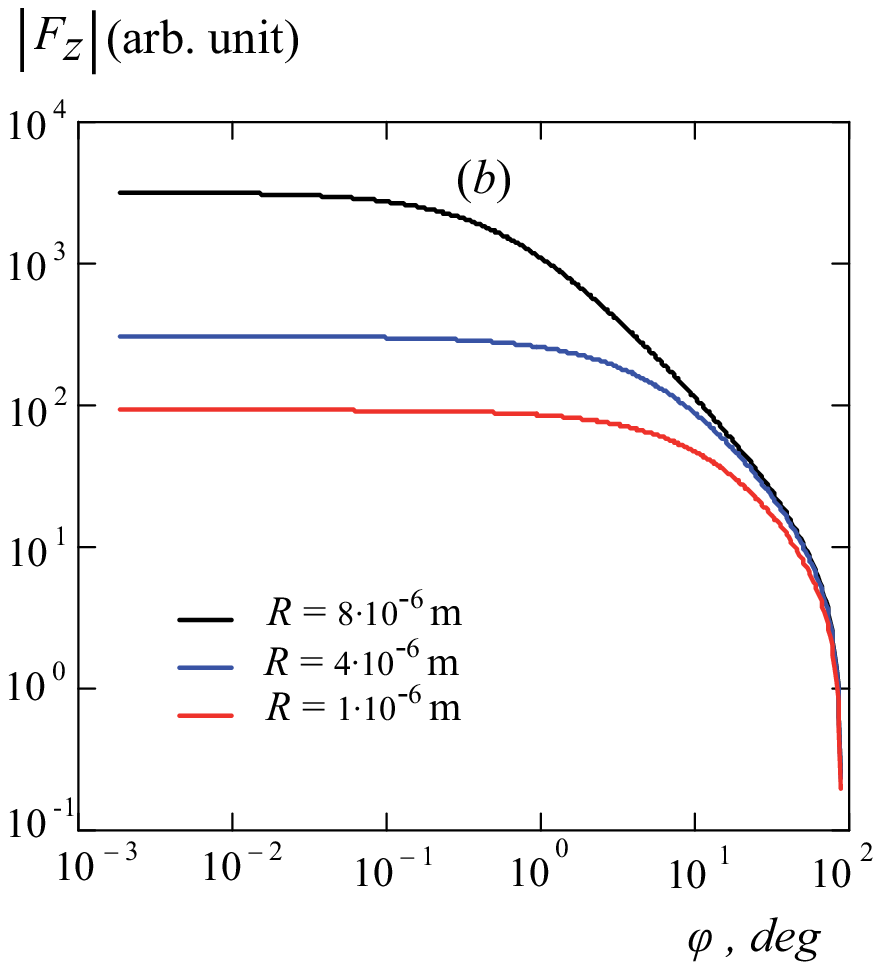}
 \end{center}
\caption{Absolute values of the Casimir forces of expulsion ($a$) and 
compression ($b$) for different lengths $R$ depending on the angle $\varphi $. }
\end{figure}

Fig. \hyperlink{fig5}{5$a$} show the dependence of total forces of expulsion and 
compression on the angle $\varphi $. It is seen that for any length of the 
surfaces of the cavity there is an expulsion maximum depending on the 
angle $\varphi $, and the larger is the length $R$, the smaller is the angle. 

The existence of noncompensated forces of expulsion is possible for any 
other open metal cavities. However, trapezoid cavities have the most diverse 
and effective properties for study and application. 

Thus, here the possibility in principle is shown that noncompensated Casimir 
force of expulsion can exist in configurations in the form of open 
(perfectly conducting) metal nanosized cavities. The force is capable of 
creating constant directed thrust which does not require the presence of 
bodies-partners. The force essentially differs from forces of repulsion 
capable of creating only effects such as levitation over bodies-partners. It 
is shown that for a trapezoid configuration, forces of expulsion have 
optimums both for sizes and angles between cavity surfaces at which their 
manifestation is maximal. A particular case of trapezoid cavities is 
parallel mirrors. It is shown here that among many other things parallel 
mirrors experience both transverse and longitudinal compression due to the 
oppositely-directed expulsion forces acting upon the ends of the mirrors. 
However for strictly parallel mirrors all the forces of expulsion are 
compensated in all directions.
\begin{acknowledgments}
The author is grateful to Yu. Prokhorov and T. Bakitskaya for his helpful
participation in discussions.
\end{acknowledgments}

\end{document}